\documentclass[useAMS,usenatbib]{mn2e}
\usepackage{mathtext,amssymb}
\usepackage{epsfig}
\usepackage{graphics}
\usepackage{graphicx}
\usepackage{url}

\newcommand{\kms}{\ensuremath{\mbox{km~s}^{-1}}}
\renewcommand{\arcsec}[1]{^{\prime\prime}\!\!\!#1\,}   
\renewcommand{\arcmin}[1]{^{\prime}\!\!\!#1\,}

\newcommand{\kpc}{\ensuremath{\rm kpc}}
\newcommand{\magdot}[1]{^{\rm m}\!\!\!#1\,}
\renewcommand{\mag}[1]{^{\rm m}\!\!\!#1\,}

\newcommand{\Lsun}{\ensuremath{\rm L_\odot}}

\newcommand{\href}[1]{\url{#1}}

\title[Modeling the optical spectrum of Romano's star]
      {Modeling the optical spectrum of Romano's star
\thanks{Based in part on data collected at Subaru Telescope and obtained
        from the SMOKA, which is operated by the Astronomy Data Center,
        National Astronomical Observatory of Japan, while other 
        data were taken from the archive of 
        Special Astrophysical Observatory (SAO) of Russian
        Academy of Sciences (RAS)}} 

\author[Maryeva and Abolmasov]{O.\ Maryeva $^{1}$\thanks{E-mail:
olga.maryeva@gmail.com} and P.\ Abolmasov$^{2}$ \\ 
$^{1}$Stavropol State University, Faculty of Physics and Mathematics, Pushkina str., 1, Stavropol, 355009, Russia \\
$^{2}$Sternberg Astronomical Institute, Moscow State University, Universitetsky pr., 13, Moscow, 119992, Russia}

\begin{document}

\date{Accepted 2011 September 9. Received 2011 August 26 ; in original form 2011 January 24}

\pagerange{\pageref{firstpage}--\pageref{lastpage}} \pubyear{2011}
\maketitle

\label{firstpage}

\begin{abstract}
          We consider the luminous blue variable (LBV) star V532
          in M33, also known as Romano's star, in two different spectral states:
          in the optical minimum of 2007/2008 and during a local brightening 
          in 2005. Optical spectra of low and moderate resolution are modeled
          using the non-LTE model atmosphere code {\sc cmfgen}.  All the
          observed properties of the object in the minimum are well 
          described by a late nitrogen-sequence Wolf-Rayet (WN) 
          star model with relatively high hydrogen
          abundance (H/He$=1.9$), 
          while the spectrum during the outburst corresponds to the spectral 
          class WN11 and is similar to the spectrum of
          P~Cyg. The atmosphere is enriched in nitrogen by about a
          factor of 6 in both states. Most of the heavy element
          abundances are consistent with the chemical composition of
          M33. Bolometric luminosity is shown to
          vary between the two states by a factor of $\sim$1.5. This makes V532
          another example of an LBV  that shows variations in its 
          bolometric luminosity during an outburst. 

\end{abstract}

\section{Introduction}
            Luminous blue variables (LBVs) are rare objects of very
            high luminosity ($\sim10^6 L_{\odot}$) and mass loss
            rates  (typically $10^{-5}{M}_\odot\mbox{yr}^{-1}$, up to
            $10^{-4} {M}_\odot\mbox{yr}^{-1}$), exhibiting strong
            irregular photometric and spectral variability  \citep{Conti,HumphreysDavidson}.  
            They are generally believed 
            to be a relatively short evolutionary stage in the 
            life of a massive star, marking the transition from 
            the Main Sequence toward Wolf-Rayet (WR) stars \citep{HumphreysDavidson,smithprinja,maeder}.  
            Significant part of the stellar mass is lost during this 
            stage forming massive circumstellar nebulae \citep{Nota95,Weis} like the spectacular 
            Homunculus surrounding $\eta$~Car \citep{smithetacar}.
            However, recent investigations of the light curves of a few 
            supernovae (SN2006gy, 2006tf, SN2005gl)  
            indicate that their progenitors underwent LBV-like eruptions 
            \citep{SmithMcCray,smith2008,GalYam2009}. 
            These observations support the view that at least some luminous LBV
            stars are the end point of the evolution  and not a transition phase, 
            but which is still in contradiction to current evolution models \citep{Meynet}.

            LBVs span quite a large range of magnitudes and
            variability types \citep{HumphreysDavidson}. 
            Some of them show  ``normal eruptions'' also called an S~Dor variability phase.
            During this phase the star may have brightness of up to 2 mag,
            but the bolometric luminosity remains essentially constant.
            Other LBVs occasionally experience ``giant eruptions'', during which
            they increase their luminosity and can reach $M \approx -14$mag (4-6 mag
            brighter than their typical quiescence magnitudes).
            Of the 35 Galactic LBVs (confirmed and candidate) \citep{clark} 
            only three ($\eta$~Car, P~Cyg and AFGL2298) 
            were observed during giant eruptions
            \citep{HumphreysDavidson,Clark09}. 
            Several examples of eruptions accompanied with changes in bolometric luminosity
            have been studied spectrally (such as HD5980 \citep{koen04}, 
            NGC~2363-V1 \citep{drissen}, AFGL2298  \citep{Clark09}).  
           These giant eruptions are not only spectacular events but are probably
           responsible for bulk of the mass loss by very massive stars
           \citep{SO06}.

            Observations of Galactic LBV stars are inevitably connected with 
            difficulties in determination of the distance and interstellar extinction, 
            that results in huge uncertainties in their bolometric luminosities. 
            Hence, studying these rare objects in nearby galaxies is 
            potentially more prospective, though extragalactic objects are more 
            distant and hence more difficult targets for spectral observations. 
            
\begin{figure}
\includegraphics[width=1\columnwidth]{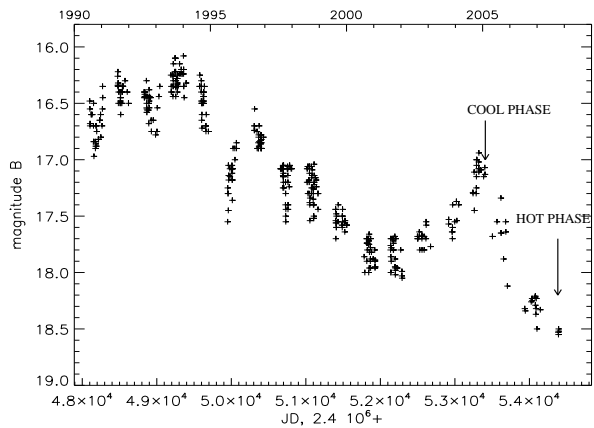}
\caption{The B-band light curve of V532. Arrows indicate the dates of
  spectral observations. 
}
\label{fig:lightcurve}
\end{figure}

            Romano's star is located  in the outer spiral arm of 
            the M33 galaxy at a distance of about $17\arcmin{\,}$\ from the centre.
            The first light curve for V532 was presented by \citet{romano}, 
            while the first spectral observations were obtained in 1992 \citep{szeifert}. 
            Because V532 demonstrates  pronounced photometrical and spectral variability \citep{kurtev,polcaro}, 
            it is classified as an LBV. The object changes from a B emission line 
            supergiant in the optical maximum \citep{szeifert}, through Ofpe/WN (WN10,WN11)
            and WN9 toward a WN8-like spectrum in deep minima \citep{me}. 
            Figure~\ref{fig:lightcurve} shows a B-band light curve of
            V532 from 1990 to 2008.

           During 2004 and 2005, V532 becomes brighter by about 1$\mag$ \ in B
           and reaches $17\mag{\,}$ in this band (see \citet{ZGphoto},
           \citet{me}, and \citet{polcaro10} for details).
           We classify the spectra obtained during this period as WN11, in
           agreement with the estimates by \citet{polcaro10}.
           Starting from the middle 2005, Romano's star weakens in all bands. 
           Its visible magnitude reaches 18$\magdot{\ .}$68 in the V band
           in February 2008.
           Recently, it exhibits a slight brightening by about 0$\magdot{\ .}$2 \citep{polcaro10}. 
           We estimated the spectral class of early 2008 to be WN8 \citep{me}. 

           In this work we investigate the optical spectra of V532 in
           two different states,
           the brightness minimum of 2008 and a moderate
           brightening in 2005, using the non-local thermodynamic equilibrium (non-LTE) radiative transfer
           code {\sc cmfgen} \citep{Hillier5}. We refer to both states as hot and cool phases, respectively.

           This paper is organized as follows. Observational data
           and data reduction process are described in Section~\ref{sec:obs}.
           In Section~\ref{sec:model} we describe the basic properties of the
           {\sc cmfgen} code. We devote Section~\ref{sec:diagdrams} to characteristic diagrams, 
           while in Section~\ref{sec:hot} 
           and Section~\ref{sec:cool} we present and analyse the modeling
           results for hot and cool phases of V532. 
           In Section~\ref{sec:disc} we discuss the results.
           Finally, Section~\ref{sec:con} we summarize the main points of our work.

\section{Observations and Data Reduction}\label{sec:obs}

              In this work, we use one spectrum of Romano's star
              obtained during the outburst in 2005   
              at the Special Astrophysical Observatory (SAO) 6-m 
              telescope~\footnote{Spectral data were taken from the archive of 
              Special Astrophysical Observatory (SAO) of Russian
              Academy of Sciences (RAS) (\href{http://www.sao.ru})}. 
              For studying  Romano's star at minimum brightness we use 
              one spectrum obtained  in January 2008 at 
              the 6-m SAO telescope and one spectrum obtained at the 
              SUBARU telescope~\footnote{Spectral data from  the SUBARU 
              telescope were taken from the SMOKA science archive \citep{smoka}
              (\href{http://smoka.nao.ac.jp})}. 
              The 6-m telescope data were obtained with the Spectral Camera with Optical
              Reducer for Photometric and Interferometric (SCORPIO)   
              in the long-slit mode \citep{scorpio}.
              One exposure 1200~s in length was obtained with the SUBARU telescope  
              with the Faint Object Camera and Spectrograph (FOCAS) \citep{focas} in October 2007. 
              VPHG450 grism was used providing the spectral range of 3750-5250~\AA. 
              Slit width of 0$\arcsec{.}5$ implies spectral resolution of about 1.7~\AA\ . 
              Observational log information on the data used in this work is summarized in 
              Table~\ref{tab:obstabscor}.

              All the  SCORPIO spectra were reduced using the {\tt ScoRe} package. 
              ScoRe was written by Maryeva and Abolmasov in IDL
              language for SCORPIO long-slit data reduction. 
              This package consists of procedures written by V.Afanasiev, 
              A.Moiseev, P.Abolmasov and O.Maryeva.
              Package includes all the standard stages of long-slit
              data reduction process. 
              FOCAS data were reduced in IDL development  environment 
              using procedures similar
              to those consisting {\tt ScoRe } but taking into account the specific
              features of FOCAS. We describe the observational data and data
              reduction process in more detail in  \citet{me}.

\begin{table*}\centering
\caption{Observational log for the spectral data used in the
  work.   S/N is signal-to-noise ratio per resolution element.} 
\label{tab:obstabscor}
\bigskip
\begin{tabular}{lcccccccc}
\hline
          &          &Exposure   &       & Spectral      &                  &          &                  & Spectral  \\ 
 ~~~~~Date&          &  time     & Grism &  range        & $\delta \lambda$ &   S/N    &   Seeing         & standard \\
          &          &   [s]     &       &    [\AA]      &    ~[\AA]        &          &[$\arcsec{\,}$~]  & star     \\
\hline
\multicolumn{9}{c}{ }\\
06.02.2005  &   SCORPIO    & $2\times 300$   &  VPHG550G& 3700-7200   &     10             &   8  & 1.7               & G248       \\
\multicolumn{9}{c}{ }\\
08.01.2008  &   SCORPIO    & $2\times 900$   & VPHG1200R&  5700-7500  & 5                  &  20  & 2.1               &  BD25d4655 \\
08.10.2007  &   FOCAS      & $1200$          &  VPHG450 &  3750-5250  & 1.7                &  30  & 0.53              &  BD40d4032 \\
\hline
\hline
\end{tabular}
\end{table*}

\section{The Model}\label{sec:model}

\subsection{CMFGEN code}
              For our analysis we used the non-LTE radiative transfer code
              {\sc cmfgen}  \citep{Hillier5}. {\sc cmfgen} has been applied to several classes of 
              objects where non-LTE effects and stellar wind are important 
              (e.g., WR, LBV, and O stars). More recently, {\sc cmfgen} was used to
              investigate the photospheric phase of Type II supernovae \citep{dessart}.
              {\sc cmfgen} solves radiative transfer equation for objects with 
              spherically-symmetric extended outflows using either the Sobolev 
              approximation or the full comoving-frame solution of the
              radiative transfer equation. To facilitate simultaneous 
              solution of the transfer equation  and statistical 
              equilibrium equations, partial linearization method is used. 
              To facilitate the inclusion of metal line blanketing in {\sc cmfgen}, 
              superlevel approach \citep{Anderson91,Anderson89} is used. 
              In this formalism, levels with similar 
              properties are treated as one and have identical
              departure coefficients. This allows to
              save a considerable amount of computer memory and time. 
              Recent versions of the code incorporate also the effect of level dissolution, 
              influence of resonances on the photoionization cross section, 
              and the effect of Auger ionization.
         
              Clumping is incorporated into {\sc cmfgen} using volume 
              filling factor approach \citep{Hillier99}. Filling factor 
              is allowed to depend on radius. By default, the wind is 
              considered homogeneous at the hydrostatic radius and becomes
              more and more clumped with the wind velocity. 
              Taking clumping into account decreases the derived mass loss rates  by
              a factor of $\sim 3-5$. The unclumped and clumped mass 
              loss rates are related to the volume filling factor $f$
              as $\dot{M}_{uncl}=\dot{M}_{cl}/\sqrt{f}$. 

              Each model is defined by the hydrostatic stellar radius $R_*$,
              luminosity $L_*$, mass-loss rate $\dot{M}_{cl}$, filling factor $f$,  wind 
              terminal velocity $v_\infty$, stellar mass $M$, and
              by the abundances $Z_i$ of included species. 
              Hydrodynamic equations are not solved, instead we propose to use 
              $\beta$-law for velocity and fix mass-loss rate for each model. This
              allows to define the density structure throughout the wind.

\subsection{Diagnostic Diagrams}\label{sec:diagdrams}

           In order to reproduce the spectrum of V532 
           obtained in October 2007 while the object was in a deep minimum 
           we took the model of the WN8 star WR40 (HD96548) calculated by \citet{HHH} and 
           gradually changed the parameters of the model.
           We calculated a grid of about 130 models with different 
           parameter values (luminosity, mass-loss rate, mass, elementary
           abundances). 
           Luminosity was varied in the range $(0.4\to 2) \times 10^6 L_\odot$. 
           For every model, we recalculated the model flux for the distance of M33.
           For our calculations we adopted a distance to M33 
           of $D=847\pm60 \ \kpc$, which gives a distance modulus 
           of  $(m-M)=24.64\pm 0.15\magdot{\,}$ \citep{distance}. 
           Then, the simulated spectra were convolved with B- and V-band
           sensitivity filters. The resulting fluxes were converted 
           to magnitudes \citep{leng} and compared to the photometrical 
           data (${\it B}=18\mag{\ .}5 \pm 0.05$ and ${\it V}=18\mag{\ .}68\pm 0.05$).  
           According to the detailed dust maps of M33 by \citet{Hippelein} 
           the effect of the intrinsic extinction should be negligible 
           near the object. Extinction in the
           Galaxy is estimated as ${\it E(B - V)\/} = 0\magdot{\ .}052$ by the NED
           extinction calculator \citep{schlegel98}. This value is well
           inside the errors for the observed ${\it B - V\/}$ colour index but
           affects the observed luminosity values that should be thus increased
           by $\sim$ 0.17~mag. 

           Mass loss rate was varied in the limits $(1.2\to 7.7)\times
           10^{-5}\rm {M}_\odot\mbox{yr}^{-1}$ typical for WN8-WN11 
           stars \citep{CrowtherWR,HamannPoWR}. 
           In all the models we assume that the volume filling 
           factor at infinity equals $f_\infty=0.1$. 
           Since V532 resides in low-metallicity environment, we considered 
           only sub-solar iron abundances (Fe/Fe$_{\odot}=0.4\pm0.15$).
           We also expected sub-solar abundances for Si, Mg, Na.

            The velocity law used was assumed a simple $\beta$-law  
            with $\beta=1$. Photospheric velocity
            was set to 100~km/s and the terminal velocity is 400~km/s 
            for all the models used for diagnostic diagrams.  
            Below we investigate the effect of variations in $\beta$ parameter
            and photospheric velocity on model spectra.
            Profile fitting for the triplet lines of He\,{\sevensize I} 
            (such as $\lambda3889,4025,4471$) allows to estimate 
            the terminal velocity as $\sim 400~~\kms$ \citep{me}.  

            Each model was classified using the equivalent width (EW) ratio of emission
            components of He\,{\sevensize I}$ \lambda 5876$ and He\,{\sevensize II}$ \lambda 5411$
            \citep{shara}. More precise estimates of physical
            conditions in the wind and atmosphere may be made using EW ratios
            which weakly depend on spectral resolution and elementary
            abundances and also allow to exclude the contribution of a possible
            circumstellar nebula.  

\begin{figure*}
\includegraphics[height =0.9\textwidth,angle=90]{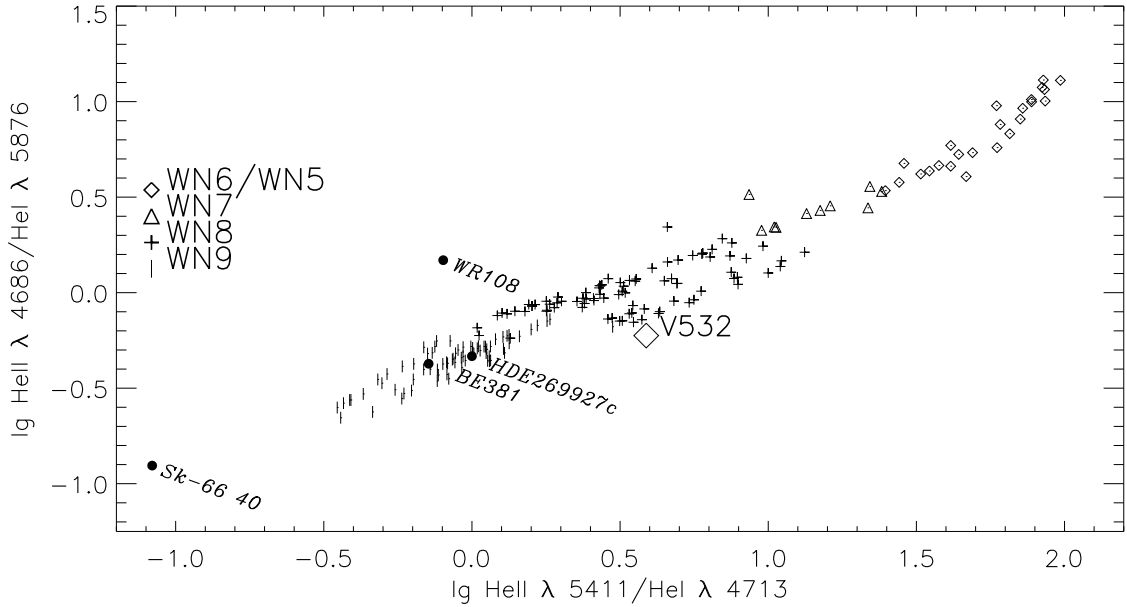}
\caption{The plot of the EW ratio logarithm log(He\,{\sevensize II} 4686 /He\,{\sevensize I} 5876) versus
                               log(He\,{\sevensize II} 5411 /He\,{\sevensize I} 4713). V532 location
                               in October 2007 (hot phase) is marked 
                               by a diamond. Sk-66 40 (WN10), BE381 (WN9), HDE269927c (WN9) 
                               and WR108 (WN9-abs) are shown for comparison. Data on these  
                               objects were taken from
                               \citet{Crowther1995}.}
\label{fig:diagram1}
\end{figure*}

            We construct several characteristic diagrams to compare 
            the model spectra with the observations using
            equivalent width ratios. Figure~\ref{fig:diagram1} shows one such diagram. 
            Modeling is complicated by the nebula surrounding V532. 
            Therefore we use the characteristic EW ratios of He\,
            {\sevensize II}$\lambda$4686 to He\,{\sevensize I}$\lambda$5876 
            (these lines form both in the stellar atmosphere and in the nebula)
            and He\,{\sevensize II}$\lambda$5411/He\,{\sevensize I}$\lambda$4713 
            (contribution of the nebula to these two lines should be negligibly small).  
            In this diagram, models fall along a narrow curve with the
            spectral class changing smoothly along it. Effective
            temperature changes from $T_*\lesssim25$kK for the lower left
            corner toward $T_*\gtrsim40$kK for the upper right.
            Model distribution in a narrow locus along a single curve is probably
            connected to the correlated behavior of emission lines of He\,{\sevensize II} as well
            as He\,{\sevensize I}. Most models shown in the diagram have identical wind velocity
            and structure but different radii, luminosities and mass loss
            rates. We see that some real stars lie near this locus 
            while WR~108 is offset. WR~108 is an unusual WN9 star --- it has 
            a higher terminal velocity ($\sim1200\mbox{km/s}$ 
            \citet{HamannPoWR}) than other WN9 stars. 
           \citet{CrowtherOfpe} classified WR~108 as intermediate 
            star between normal Of and WN stars. 

            Bright hydrogen lines are present in the spectra,
            evidently stronger than in the spectra of ordinary late WN
            stars. In this sense, Romano's star is similar to H-rich WN stars. 
            Therefore we calculate models with   
            H/He$ =0.75\to 2.6 $ (by number) that is typical for hydrogen WR stars. 
            For the models that reasonably fit the observed equivalent
            widths, we varied the chemical composition and filling
            factor to fit the line strengths of individual elements
            and line profiles, correspondingly.

\subsection{Hot-Phase Spectrum}\label{sec:hot}

\begin{figure*}
\includegraphics[width =1\textwidth]{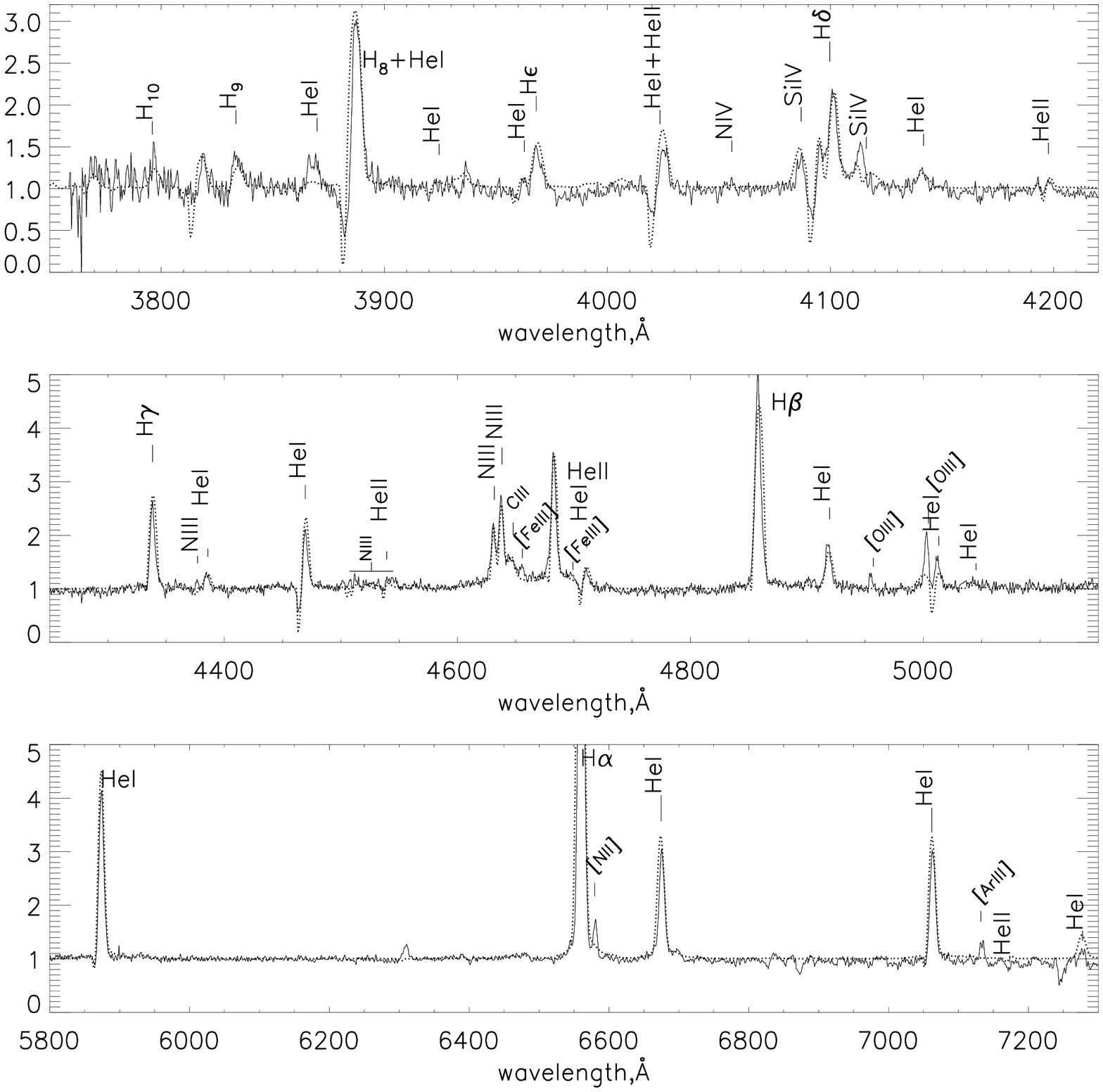}
\caption{Normalized hot-phase optical spectra compared with the best-fit {\sc cmfgen} model (dotted line). 
         Bottom panel shows the spectrum obtained with SCORPIO, top
         and middle panels show the spectrum obtained with
         FOCAS. The model spectrum on bottom panel is convolved 
         with the  5\AA-wide Gaussian instrumental profile.} 
\label{fig:hotmodel}
\end{figure*}

           In Figure~\ref{fig:hotmodel} we show the observed 
           spectra of V532 at minimum brightness and the best-fit model spectrum.  
           Figure~\ref{fig:hotmodel} shows good agreement between the observed and
           model spectra. Both in observational data and in the best-fit model,
           singlet and triplet lines of He\,{\sevensize I} show different profile shapes:
           triplet lines have typical P~Cyg profiles while singlets are pure
           flat-topped emissions.

           For this model, the best-fit volume filling factor at infinity
           $f_\infty=0.1$, mass loss rate $\dot{M}_{cl}=1.9\times          
           10^{-5}\rm M_{\odot} yr^{-1}$ (which correspond to unclumped
           $\dot{M}_{uncl}\simeq 6\times 10^{-5}\rm M_{\odot} yr^{-1}$). 
           The stellar temperature $T_*$ follows from the relation 
           $L_*=4\pi R^2_* \sigma T_*^4$, and the effective temperature
           at the photosphere $T_{eff}$ is defined by the Rosseland optical depth of $2/3$. 
           The values obtained for $T_*$, $T_{eff}$ and other physical 
           parameters are listed in Tables~\ref{tab:parmodelbeta}
           and~\ref{tab:parmodel}. 
           Errors in luminosity given in the table are determined by the
           uncertainties of the photometrical data.
           The errors in the H/He ratio are from the model fitting alone. 

\begin{table*}\centering
\caption{Best-fit model parameters for different values of $\beta$ and
  photospheric velocity. 
H/He denotes hydrogen number fraction relative to helium.} 
\label{tab:parmodelbeta}
\bigskip
\begin{tabular}{lccccccccc}
\hline
 $\beta$ &$V_{phot}$& $V_{\infty}$& $T_*$     &   $R_*$        & $T_{eff}$ &   $R_{2/3}$    &     $L_* $    & $\dot{M}_{cl}$                 & H/He  \\
        &  [km/s]   &   [km/s]    &[kK]       &[$\rm R_{\odot}$]&  [kK]      &[$\rm R_{\odot}$]&[$10^5~L_{\odot}$]&[ $10^{-5} \rm M_{\odot}\,yr^{-1}$]&  \\
\hline
1        &  100  &  400   &$34.0\pm0.5$&$20.8\pm1.$     &$31.7\pm0.5$&  $23.9\pm0.7$ &   $5.2\pm0.2$  & $1.9\pm0.2$                     & $1.9\pm0.2 $  \\
         &       &        &            &                &            &               &                &                                 &              \\
1        &  10   &   400     &$33.0\pm1.0$&$21.6\pm1.2$    &$29.4\pm0.5$&$27.3\pm2.0$   & $5.0\pm0.25$   & $1.6\pm0.1$                    & $1.9\pm0.2 $ \\ 
         &        &       &            &                &            &               &                &                                 &     \\ 
4        &  100   &   500   &$33.0\pm1.0$&$21.6\pm1.2$    &$29.0\pm1.0$& $27.8\pm2.0$  &   $5.0\pm0.25$ & $1.6\pm0.15$                    &  $2.0\pm0.2 $ \\ 
         &         &      &            &                &            &               &                &                                 &  \\
\hline
\end{tabular}
\end{table*}

            Surface chemical abundances obtained for V532 are listed in Table~\ref{tab:frac}.
            Abundance pattern is consistent with the moderately sub-solar metallicity 
            of M33 ([Fe/H]$ \sim -0.5$), but nitrogen is significantly over-abundant 
            ($\sim 6.4 Z_{\odot}$ ). The latter value is consistent with the existing
            evolutionary models and with the data on other nitrogen-rich WR stars \citep{HHH}.
            Silicon abundance was adjusted using Si\,{\sevensize III}~$\lambda 4565.05$ and
            Si\,{\sevensize IV}~$\lambda\lambda 4088.90, 4116.10$ lines, magnesium abundance using
            Mg\,{\sevensize II}~$\lambda 4481.13$.  Ne, Al, S, Ar and Ca abundances were
            fixed relative to He because they are poorly constrained by
            observational data due to the lack of strong lines for these
            elements. We assume these abundances identical
            to those for WR40. 
\begin{table}\centering
\caption{Derived stellar abundances for Romano`s star. 
         Abundances for all the elements 
         are identical for the two phases within the uncertainties. 
}
\label{tab:frac}
\bigskip
\begin{tabular}{lccc}
\hline
SPECIES  &   Number Fraction  &   Mass Fraction &   Z$_i$/Z$_\odot$  \\
\hline
     H     &    $1.9$                 &    $3.18\times 10^{-1}$ &  $0.45 $  \\ 
     He    &    $1.0$                 &    $6.7\times 10^{-1}$  &  $2.4$   \\    
     C     &    $1.0\times 10^{-4}$   &    $2  \times 10^{-4}$  &  $0.07$   \\   
     N     &    $3.0\times 10^{-3} $  &    $7  \times 10^{-3}$  &  $6.4 $   \\  
     O     &    $8.0\times 10^{-4}$   &    $2.2\times 10^{-3}$  &  $0.23$   \\  
     Ne    &    $2.4\times 10^{-4}$   &    $8.1\times 10^{-4}$  &  $0.47$  \\ 
     Na    &    $7.0\times 10^{-6}$   &    $2.7\times 10^{-5}$  &  $0.78$    \\ 
     Mg    &    $5.0\times 10^{-5}$   &    $2  \times 10^{-4}$  &  $0.31$   \\
     Al    &    $8.4\times 10^{-6}$   &    $3.8\times 10^{-5}$  &  $0.7 $   \\    
     Si    &    $2.0\times 10^{-4}$   &    $1.0\times 10^{-3}$  &  $1.43$    \\   
     S     &    $2.3\times 10^{-5} $  &    $1.2\times 10^{-4}$  &  $0.34$   \\     
     Ar    &    $6.5\times 10^{-6}$   &    $4.4\times 10^{-5}$  &  $0.43$  \\   
     Ca    &    $4.2\times 10^{-6}$   &    $2.8\times 10^{-5}$  & $0.44$  \\  
     Fe    &    $4.87\times 10^{-5}$  &    $4.6\times 10^{-4}$  & $ 0.34$  \\  

\hline   
\end{tabular}
\end{table}

            After we obtained a model reasonably approximating the observed
            spectra and consistent with the  photometrical data we 
            investigated the effects of photospheric velocity on the model spectrum and 
            the derived model parameters. 
            Decrease in photospheric velocity implies increase in Rosseland optical depth 
            that affects both $T_{eff}$ and $R_{2/3}$.            
            These changes practically do not affect hydrogen and neutral helium lines, 
            while the EW of the He\,{\sevensize II}$\lambda$4686 emission increases by about 14 per cent
            when photospheric velocity decreases by a factor of two. 
            Table ~\ref{tab:parmodelbeta} gives the parameters of a model with
            $v_{phot}=10~\kms$ that reproduces the spectrum indistinguishable from the
            current best-fit model with $v_{phot}=100~\kms$. One order of
            magnitude uncertainty in photospheric
            velocity results in 3 per cent and 10 per cent uncertainties in luminosity and mass
            loss rate, respectively. 

\subsubsection{Hot-Phase Spectrum with $\beta>1$}\label{sec:beta}

       As stated above the velocity law used was assumed a simple $\beta$-law: 
       $$                                                               
       v(r) \simeq v_{\infty}\left(1-\frac{r_0}{r}\right)^\beta,        
       $$                                                               
        with 
       $$                                                                               
        r_0 = R_*\left\{1-\left(\frac{v_{phot}}{v_{\infty}} \right)^{1/\beta} \right\}  
       $$                                                                               
       where $v_{phot}$ is photospheric velocity.
       Above we fixed the $\beta$ parameter to unity. 
       In some objects in certain phases, however, $\beta$ 
       attains a significantly higher value, up to $\beta \sim 4$ \citep{najarro97,groh}.
       In general, LBV stars have low $\beta \lesssim 1$ during optical
       minima, but the velocity law changes during eruptions \citep{guoli07}.
       Let us now consider the effect of variable $\beta$ parameter on model spectrum. 

\begin{figure*}
\includegraphics[width=1\textwidth]{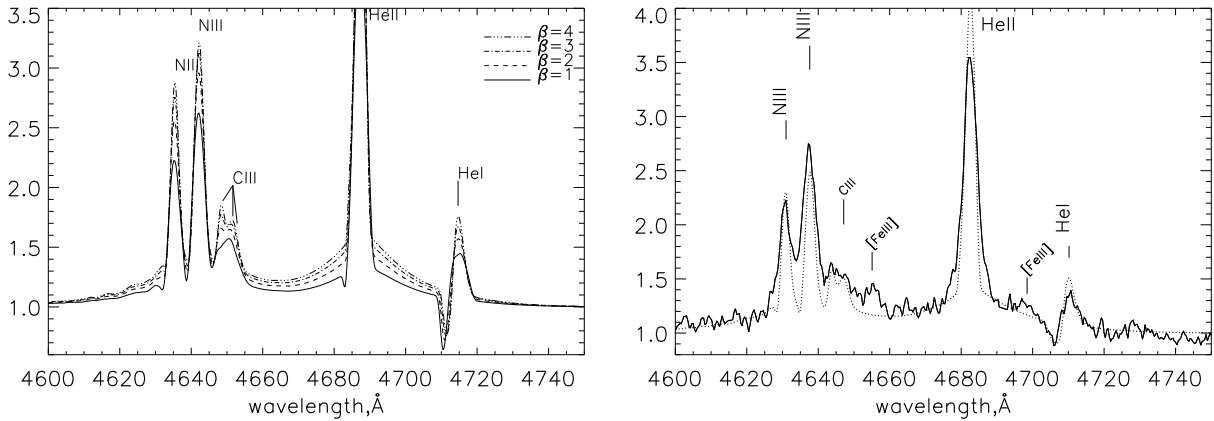}
\caption{Left-hand panel: effects of parameter $\beta$ on the profile of the 4630-4713 \AA \ blend. 
         Right-hand panel:  fitting of the 4630-4713 \AA \ blend. Observational spectrum 
         is shown by a solid line, and the dotted line shows our hot-phase model calculated with {\sc cmfgen} ($\beta=4$).  
}
\label{fig:beta}
\end{figure*}

      Figure~\ref{fig:beta} illustrates the dependence of the 4630-4713
      \AA\ blend shape on $\beta$. 
      Increasing $\beta$ produces higher equivalent widths of all the
      detectable lines of the blend. Besides, the wings of
      He\,{\sevensize II}$\lambda4686$ (see also section~\ref{sec:hewings}) become stronger. 
      As a result, intensities of C\,{\sevensize III} $\lambda\lambda4647.4,4650.3,4651.5$
      lines change, while the line ratios of C\,{\sevensize III}~$\lambda\lambda4647.4,4650.3,4651.5$ 
      remain somewhat constant. For large  $\beta$  ($\beta\ge2$) 
      C\,{\sevensize III}$\lambda 4647.4$ is stronger than other carbon lines, and the
      shape of the 4630-4713\AA\ blend changes. 
      Intensity ratios of the carbon lines observed in the spectrum of
      V532 (see the right-hand panel of Figure~\ref{fig:beta}) indicate $\beta\simeq 4$.
      An increase in $\beta$ affects the model spectrum similarly as an increase 
      in the mass-loss rate. Therefore we may describe the observed spectrum by a model 
      with $\beta=4$ and a lower (by about 10 percent) mass-loss rate.

\begin{figure*}
\includegraphics[width =1\textwidth]{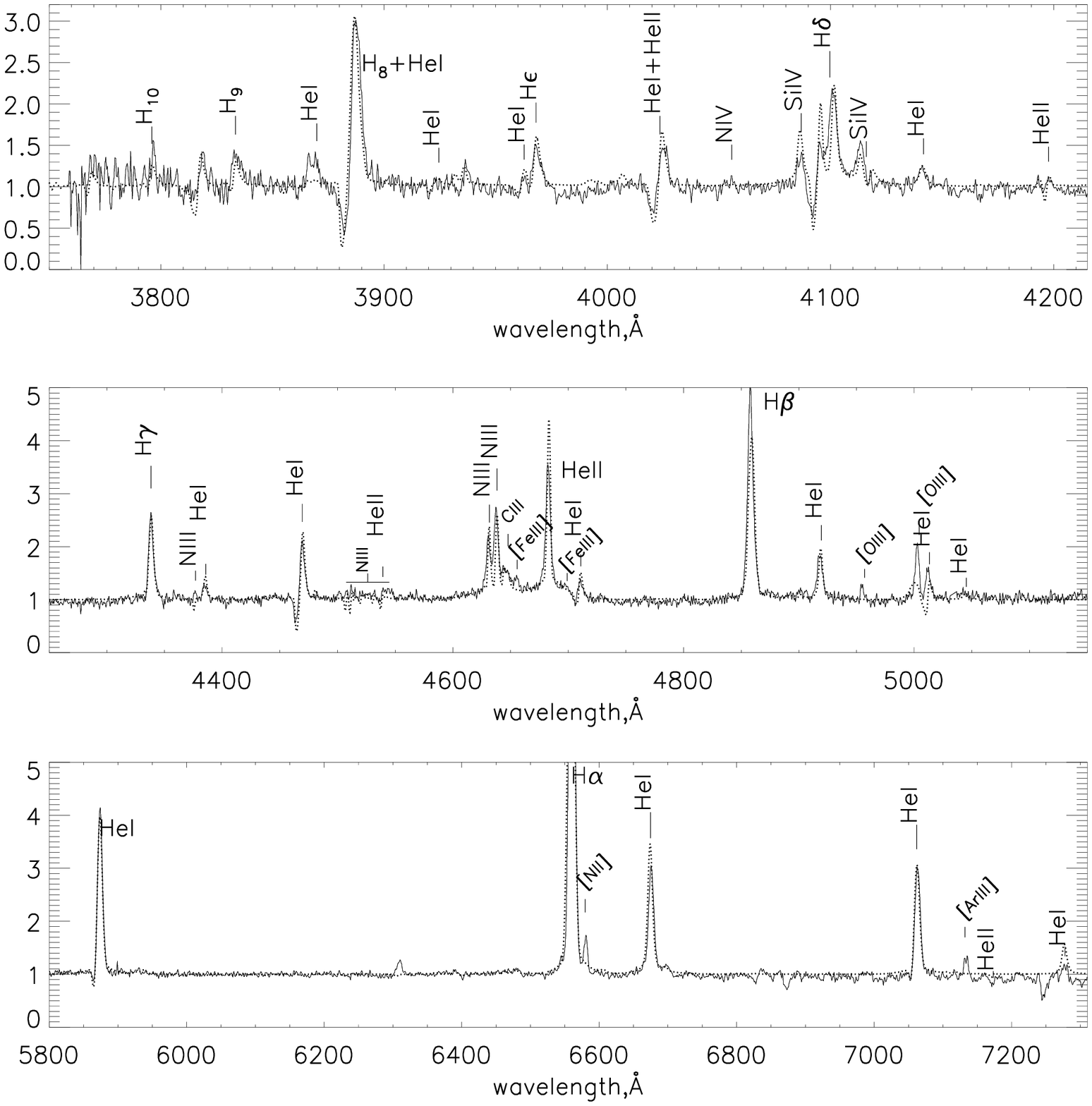}
\caption{The normalized hot-phase optical spectra compared with {\sc cmfgen} model with $\beta=4$ (dotted line). 
         Bottom panel shows the spectrum obtained with SCORPIO, and top
         and middle panels show  the spectrum obtained with
         FOCAS. The model spectrum on the bottom panel is convolved 
         with the 5\AA-wide Gaussian instrumental profile.} 
\label{fig:spectrumbeta4}
\end{figure*}

       Figure~\ref{fig:spectrumbeta4} shows 
       comparison between the observed spectrum of Romano's star and the model with $\beta=4$. 
       The model with $\beta=4$ fits better the triplet lines of helium, 
       while  He\,{\sevensize II}$\lambda4686$ is brighter than for the best-fit model with $\beta=1$.
       Table  \ref{tab:parmodelbeta} shows the best-fit parameters for
       the models with different values of photospheric velocity
       $V_{phot}$ and $\beta$. 

\subsection{Cool-Phase Spectrum}\label{sec:cool}

           During 2004 and 2005, V532 becomes brighter by about 1 mag in B
           and reaches 17 mag in this band. 
           Colour index ${\it B - V\/}$ is constant within the observational
           errors (${\it B - V\/} = -0.11$ mag for 2007  October and
           $-0\magdot{\ .}17$ for February 2005). 
           Spectra obtained during this brightening we classify as WN11, in
           agreement with the estimates of \citet{polcaro10}.
           For studying Romano's star in the cool phase, we use the spectrum 
           obtained in 2005 February  at the 6-m SAO telescope.

\begin{figure*} 
\begin{center}
\includegraphics[width=1\textwidth]{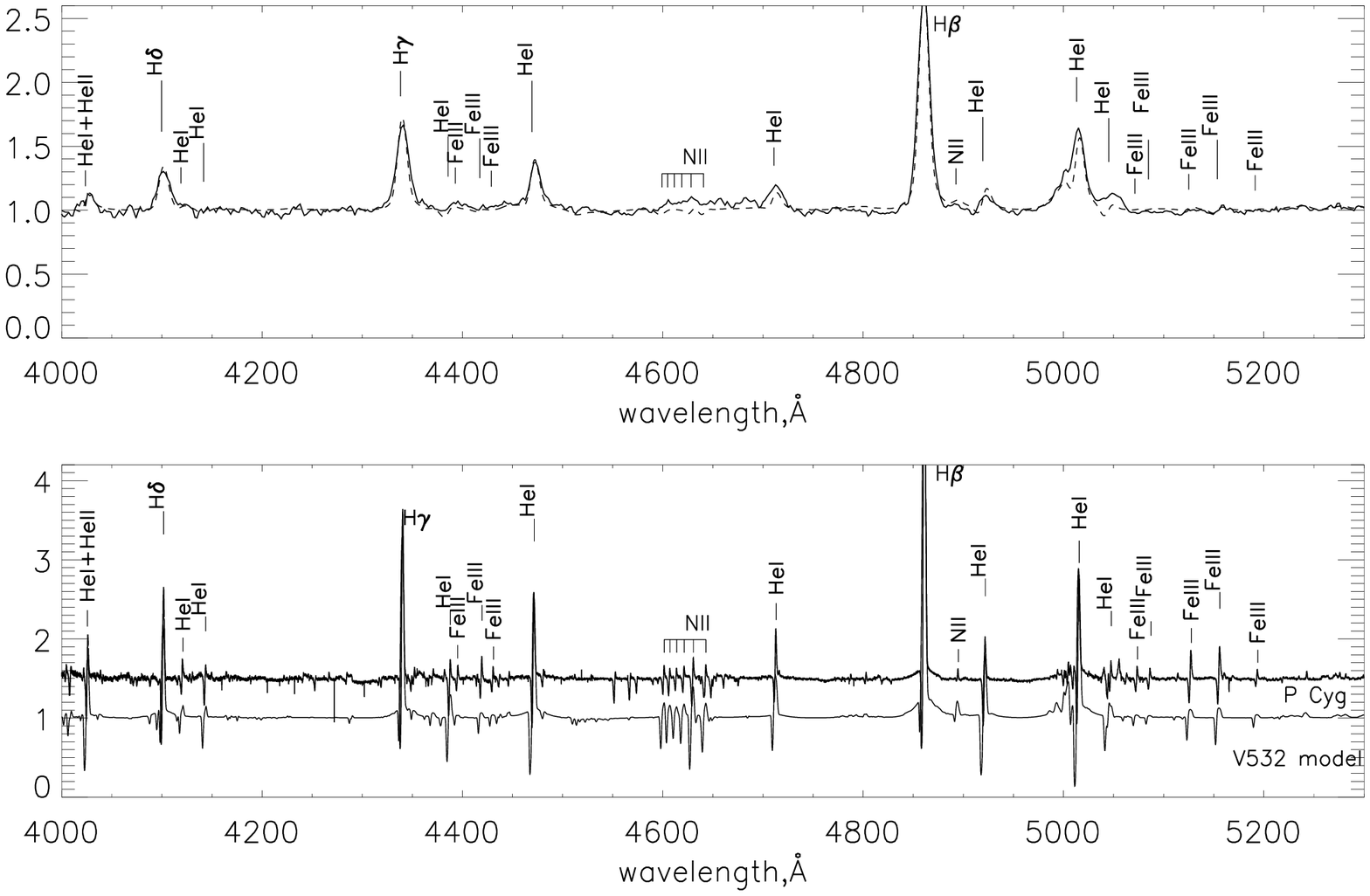}
\end{center}
\caption{Fitting the cool-phase spectrum in the blue spectral range. 
Top panel: spectrum of V532 (Feb.~2005) (solid line) and our
  cool-phase model (convolved with the instrumental profile).   
 Bottom: comparison of our cool-phase model spectrum with the archival spectra
 of the LBV star P~Cyg.
 Spectra are normalized by the local continuum level.}
\label{fig:coldmodel1}
\end{figure*}

\begin{figure*}
\begin{center}
\includegraphics[width=1\textwidth]{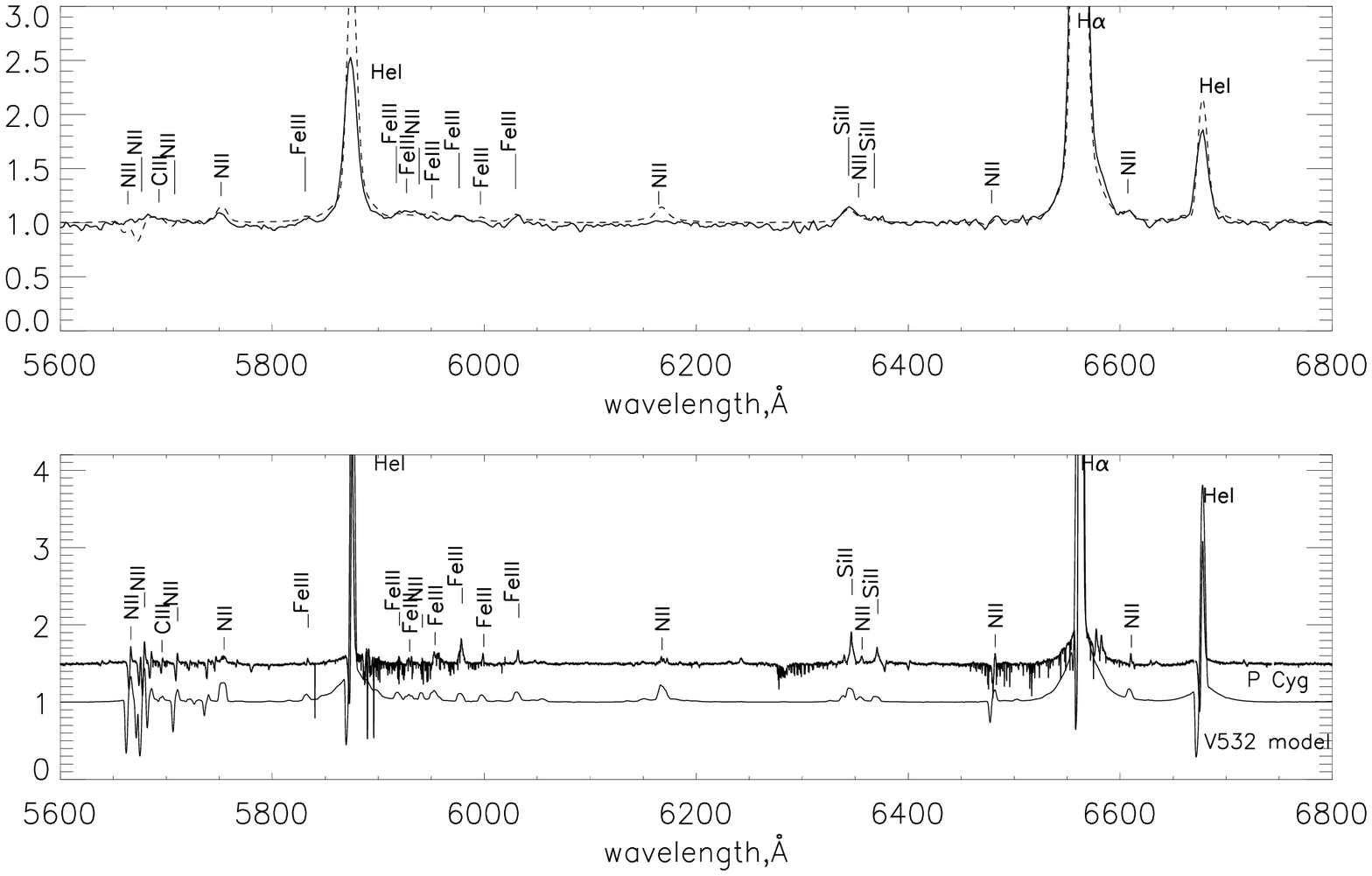}
\end{center}
\caption{Fitting the cool-phase spectrum in the red spectral range.
Top panel: Spectrum of V532 (Feb.~2005) (solid line) and our
  convolved cool-phase model.   
 Bottom: comparison of our cool-phase model with observation spectrum of LBV star  P~Cyg.
 Spectra are normalized by the local continuum level.} 
\label{fig:coldmodel2}
\end{figure*}

           Using the hot model as initial, we increased the mass loss
           rate and hydrostatic radius thereby decreasing the effective temperature. 
           After obtaining a model consistent with the photometrical 
           data (${\it V}=17.27\pm 0.03$mag, ${\it B}=17.1\pm 0.03$mag, assuming $A_v=0.17$mag) 
           we started varying 
 the wind velocity  and volume filling factor. 
            The EW  of H$\alpha$ increases by 30~ per cent when $f_\infty$ is
            increased from 0.1 to 0.5. But the effective temperature 
            at the photosphere and the photosphere radius varies insignificantly.
            Mass loss increase leads to H$\alpha$ EW increase as well. A 18 per cent
            variation of mass loss rate leads to approximately 18 per cent variation in
            the EW value. However, it also leads to significant changes in effective
            temperature and photosphere radius. Therefore we fixed $\dot{M}$ and
            changed only $f_\infty $.
           To compare the model spectrum with the observations 
           we convolved the observed spectrum with the 10\AA-wide Gaussian
           instrumental profile.
           Figures~\ref{fig:coldmodel1} and \ref{fig:coldmodel2} show  
           the observed spectrum and the best-fit model spectrum
           convolved with the instrumental profile. 
           The figures also show
           comparison between the raw (high-resolution) best-fit model spectrum and
           a spectrum of the LBV star P~Cyg (B1Ia$^+$) obtained in August
           1998 and taken from {\it Elodie } archive (\url{http://atlas.obs-hp.fr/elodie/intro.html}).
           The spectra are unexpectedly similar.

           For the best-fit model in the cool phase, volume filling factor at infinity
           $f_\infty=0.5$. 
           This value is factor of 5 higher than the
           one typically found for WR stars \citep{HHH}. Note that this value is equal to that for P~Cyg
           \citep{NajarroPCyg}, thus confirming the similarity of
           these objects. The volume-filling factor seems to depend strongly either on
           the mass loss rate or on the spectral state, changing from
           $\sim 0.1$ for late WN stars to $\sim 0.5$ for
           B-type hypergiants. Self-consistent modeling of wind
           acceleration is needed to understand the mechanisms leading
           to the strong clumpiness of the winds of Wolf-Rayet stars.

           The mass loss rate for this model is
           $\dot{M}_{cl}=(4.5\pm 0.2)\times 10^{-5}\rm M_{\odot} yr^{-1}$, 
           luminosity is $(7.7\pm0.25)\times 10^5\Lsun$, $T_{eff}=20.4\pm1.0~\mbox{kK}$. 
           Wind and stellar parameters of V532 in the  maximum of brightness are given in  Table~\ref{tab:parmodel}.  
           Surface chemical abundances are the same as for the hot-phase model 
           (see Table~\ref{tab:frac}).  

\section{Discussion}\label{sec:disc} 
        
\subsection{Evolution of the Physical Parameters}

         We model the spectra of V532 in maximum of brightness 
         ({\it V}=$17\mag{\,}$, Feb.~2005) and in minimum of brightness 
         ({\it V}=$18\mag{\ .}6$, Oct.~2005) using the non-LTE 
         radiative transfer code {\sc cmfgen}.  
         Stellar parameters derived for both hot- and cool-phase
         models are given in Table~\ref{tab:parmodel}. 
         For comparison, the values of these parameters for some other WN stars
         taken from the literature are given in the table.
         Note that the parameters of the Galactic WN8 stars WR124, WR16 and
         WR40, as well as the LBV AG~Car in photometric minima and LBV P~Cyg were calculated
         using {\sc cmfgen} models taking clumping into account (see
         \citet{Crowther99,HHH,groh,NajarroPCyg}). 
         On the other hand modeling of the WN9h stars R84 and BE381 
         did not account for clumping, and that may lead to 
         mass loss rate overestimates.
         Table~\ref{tab:parmodel} shows that in minimum 
         brightness, V532 is similar to a classical WN8 star, but the wind
         velocity is lower, characteristic instead for a WN9 star. 
         We observe that relative hydrogen abundance (H/He) for V532 is 
         similar to that of WN8h stars, as given in the table. 
         V532 as well as other LBVs AG~Car and P~Cyg (\citet{groh,NajarroPCyg}) 
         shows significant enhancement of helium (more than a factor of 2 relative to solar), 
         impying the end of the hydrogen shell-burning phase. 
         High nitrogen content and depletion of carbon and oxygen 
         are indicative of material which has undergone the CNO cycle. 
         Our results agree with the contemporary understanding of V532 and
         Galactic LBV stars as a transitional stage from main-sequence
         supergiants to WR stars. 

         During the 2005 February outburst, 
         parameters of the star 
         correspond to the spectral class WN11. 
         The model spectrum is similar to the spectrum of P~Cyg. 
         V532 shows a WN11 spectrum in the maximum, while the
         classical LBVs such as AG~Car and P~Cyg show the same spectrum only in the deep
         minima and  the long-lasting quiet state, respectively. 
         Note however that V532 had a strong maximum in 1993 ($0\magdot{\ .}9$ brighter than
         in February 2005) and exhibited a B-supergiant spectrum. 
         Hence, V532 shows stronger spectral variability than AG~Car.

         The two phases, hot and cool, are mainly distinguished by the
         photosphere radius. In the hot phase the radius is about three times larger
         in the cool phase.
         Three basic parameters vary simultaneously and make measurable
         contributions to the observed inflation of the star. 
         For the two states, $\dot{M}$ differ by a factor 2.4, and the wind
         velocity is 1.8 times larger for the hot state. It is easy to check
         that the size of the wind photosphere should scale
         approximately as:

$$
R_{ph} \propto \left(\frac{M_{\odot}}{v_\infty}\right)^n,
$$

         where $n=1$ in the case of pure scattering wind 
        (mass-absorption coefficient $\varkappa =  const$)
         and $n=2/3$ when $\varkappa \propto \rho$, as for true
         absorption processes. Hence, $R_{ph}$ is expected to vary by a factor
         $3\to 5$, in consistence with the observed value of $2.9$. 
         Our model favours correlation of  hydrostatic radius with
         mass loss rate. 

        Romano's star  is situated near a young OB
        association OB~89. Probably V532 was a member of OB~89 and 
        was ejected via slingshot-type dynamical interaction. More detail we consider this 
        suggestion in Maryeva \& Abolmasov (in preparation). 
          
\begin{table*}\centering
\caption{Derived properties of V532 in the maximum and minimum 
  brightness, and comparison with related stars in M33, Large Magellanic Cloud (LMC) and Milky
  Way (MW) galaxies, including the LBVs P~Cyg and AG~Car in visual minima (1990 December, 2002 July).
} 
\label{tab:parmodel}
\bigskip
\begin{tabular}{lcccccccccccc}
\hline
Star      & Gal. & Sp.   & $T_*$ &   $R_*$ & $T_{eff}$ &$R_{2/3}$   & $\log L_*$&$\log \dot{M}_{cl}$          & f  & $v_{\infty}$& H/He & Ref       \\
          &      & type  & [kK]  & [$\rm R_{\odot}$] &   [kK]  &[$\rm R_{\odot}$]&[$\rm L_{\odot}$]&[$\rm M_{\odot}\,yr^{-1}$]&    & [\kms]      &      &            \\
\hline
 WR124    & MW   &  WN8h & 32.7  &    18.0   &           &             &   5.53    &     -4.7         & 0.1&  710        & 0.7  & [1] \\ 
 WR40     & MW   &  WN8h & 45.0  &    10.6   &           &             &   5.61    &     -4.5         & 0.1&  840        & 0.75 & [2] \\ 
 WR16     & MW   &  WN8h & 41.7  &    12.3   &           &             &   5.68    &     -4.8         & 0.1&  650        & 1.2  & [2] \\ 
          &      &       &       &           &           &             &           &                  &    &             &      &     \\
 R 84     & LMC  & WN9h  & 28.5  &   33.8    &    24.9   &  44.2       &   5.83    & -4.40            &    &  400        & 2.5  & [4]  \\
 BE 381   & LMC  & WN9h  & 30.6  &   20.8    &    27.5   &  26.0       &   5.54    & -4.65            &    &  375        & 2.   & [4]  \\
          &      &       &       &           &           &             &           &                  &    &             &      &     \\
 AG~Car   &  MW  & WN11  & 24.64 &    67.4   &  21.5     &   88.5      &   6.17    &     -4.82        & 0.1&  300        &  2.3    &[5]  \\
Dec 1990  &      &       &       &           &           &             &           &                  &    &             &      &         \\ 
 AG~Car   &  MW  & WN11  & 18.7  &    95.5   &  16.4     &   124.2     &   6.0     &    -4.33         &0.25&  195        &      & [5] \\
Jul 2002  &      &       &       &           &           &             &           &                  &    &             &      &       \\
P Cyg     &  MW  &B1I$a^+$&      &           &  18.7     &   76.0      &   5.8     &    -4.63         & 0.5&  185        & 2.5  & [3]  \\
          &      &       &       &           &           &             &           &                  &    &             &      &      \\ 
 V532     &  M33 &  WN8  & 34.0  &  20.8     &  31.7     &  23.9       &   5.7     &    -4.72         & 0.1&  $360^*$        &  1.9 &     \\ 
hot-phase &      &       &       &           &           &             &           &                  &    &             &      &     \\ 
 V532     &  M33 &  WN11 & 22.0  &  59.6     &  20.4     &  69.1       &  5.89     &    -4.4          & 0.5&  200        &  1.4 &     \\
cool-phase&      &       &       &           &           &             &           &                  &    &             &      &      \\ 
\hline
 \multicolumn{13}{l}{\footnotesize [1]- \citet{Crowther99}, [2]- \citet{HHH}, [3]- \citet{NajarroPCyg}, [4]- \citet{CrowtherLMC}, [5] - \citet{groh}}\\
\multicolumn{13}{l}{$^{*}$ -- $v_{\infty}$ was estimated using He\,{\sevensize I} lines \citep{me}}
\end{tabular}
\end{table*}

            \citet{polcaro10} presented the values of bolometric luminosity, 
            effective temperature and radius of Romano's star, using 
            bolometric corrections for known WN8 and WN11 stars. 
            Here, we use a more comprehensive way to estimate the bolometric
            luminosity. But the main conclusion holds: bolometric luminosities of V532 were
            different in 2005 and 2008. The luminosity of V532 in 2005 
            ($L_*=7.7\cdot10^{5}L_{\odot}$) is  1.5 times higher. 
            Therefore, V532 should be considered one more LBV (after
            the objects mentioned by
            \citet{koen04,drissen,Clark09}) that changes its luminosity
            during (even moderate amplitude) eruption. 
            In this sense, V532 behaves similarly to AG~Car that has bolometric
            luminosity variations during its S~Dor cycle \citep{groh}. 

\subsection{The Broad Wings of the He\,{\sevensize II}$\lambda4686$ Emission}\label{sec:hewings}

\begin{figure}
\begin{center} 
\includegraphics[width=1.1\columnwidth]{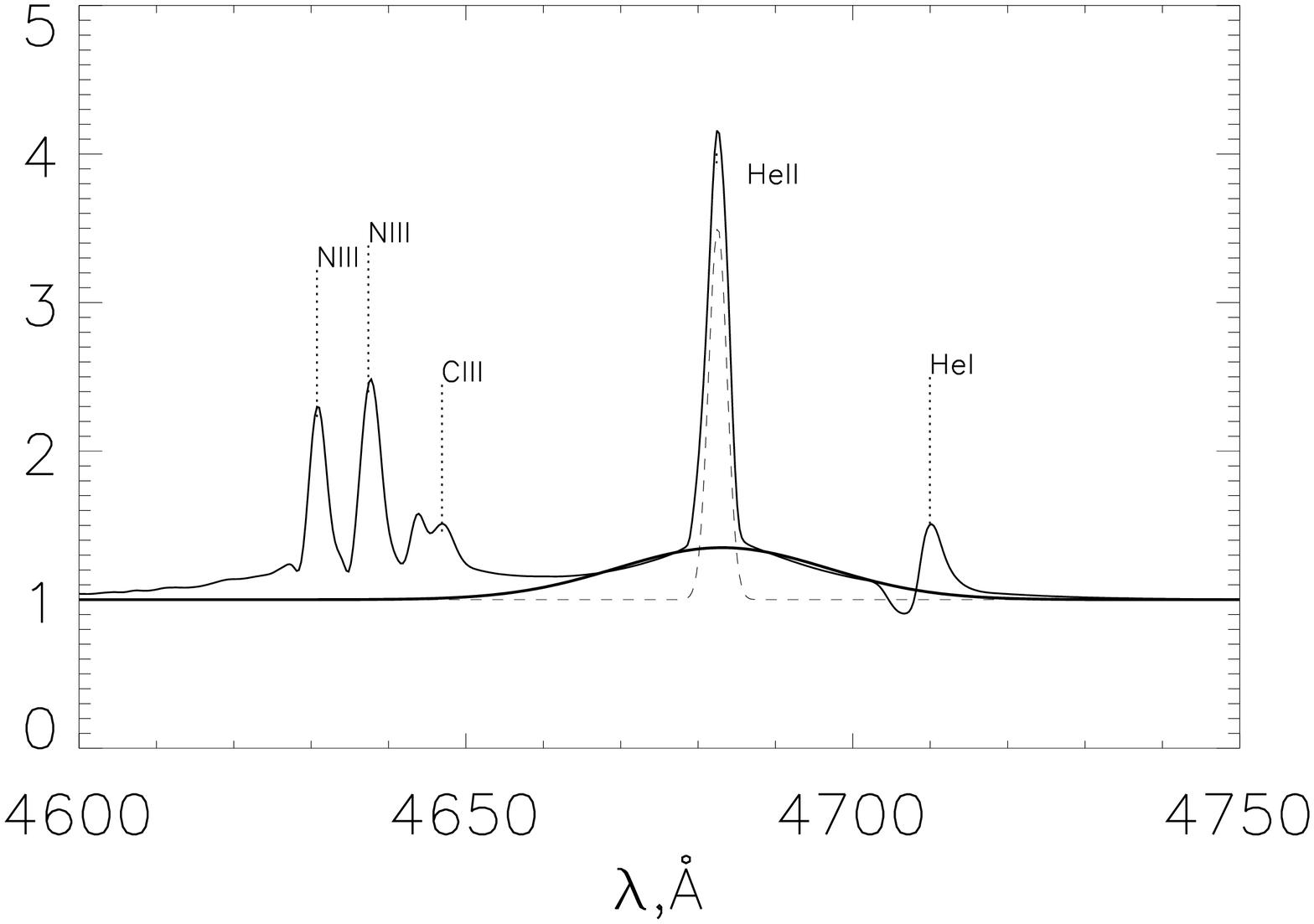}
\end{center}
\caption{Approximation of the
         He\,{\sevensize II}$\lambda 4686$ emission in the model spectrum by two Gaussians. The broad 
         component of He\,{\sevensize II}$\lambda 4686$ is shown by a thick line, 
         narrow by a dashed line.} 
\label{fig:wrbumpmodel}
\end{figure}

           \citet{polcaro10} detect a broad ($\sim 1000\,\kms$) component 
           of the He\,{\sevensize II}$\lambda4686$ line in the spectrum of V532 
           obtained in December 2008. Authors explain this component by 
           a bimodal stellar wind.
           Although {\sc cmfgen} adopts spherical symmetry, the model profiles 
           of this emission line possess similar, and sometimes even broader and brighter wings
           (Fig.~\ref{fig:wrbumpmodel}). Observed wings may 
           be explained by a more common phenomenon -- by electron scattering.
           Strong and very broad emission wings  were first found for LBVs  
           in P~Cygni by \citet{Bernat}. Balmer lines in the spectrum of 
           AG~Car do have such wings, which extend to more than $\pm 1500\,\mbox{km s}^{-1}$ 
           from the core of the line \citep{stahl2001}. For P~Cyg,
           electron scattering wings were reproduced by
           \citet{najarro97} in spherical symmetry. 
           The wings are explained by scattering 
           of line photons by free electrons of the stellar wind. Large 
           widths of the wings reflect high thermal velocities of free electrons 
           and does not correspond to bulk motion of the gas. 

\section{Conclusions}\label{sec:con}

        Using comoving frame numerical radiative transfer with the {\sc cmfgen}
        code, we estimate the physical parameters of the photosphere of
        Romano's star coming to the two principal conclusions. First, variability in
        this object is caused by correlated changes in mass-loss rate, wind velocity and  
        hydrostatic radius. Secondly, elementary abundances do not change
        significantly, we find similar helium and
        nitrogen overabundance in both states, characteristic for hydrogen-rich WNL
        stars, H/He$ \simeq 1.7-2.0$
        and N/He$ \simeq (3-5)\times 10^{-3}$. 

         We find that the bolometric luminosity of this object was higher during the
        eruption in 2005 by a factor of $\sim$1.5, that makes V532 one more example
        of an LBV that changes its luminosity. Together with the
        moderate intensity outburst of AFGL2298, its behaviour
        indicates that even moderate amplitude LBV outbursts are
        accompanied by changes in bolometric luminosity.

\section*{Acknowledgments}
        We would like to thank
        the anonymous referee for valuable comments. 
        The paper is partially based on spectral data retrieved from the ELODIE archive at
        Observatoire de Haute-Provence (OHP). 
        Based in part on data collected at Subaru Telescope, which is operated
        by the National Astronomical Observatory of Japan, and taken
        from the SMOKA, which is operated by the Astronomy Data Center,
        National Astronomical Observatory of Japan. We also use the
        data from the archive of the Special Astrophysical Observatory. 
        We also wish to thank John D. Hillier for his great code {\sc cmfgen}, both
        comprehensive and user-friendly, that we applied to fit and analyse
        the data. 
        One of us (P. A.) thanks leading
        scientific schools grant NSh-7179.2010.2 for support.


\label{lastpage}

\end{document}